\documentclass[twocolumn,aps,pra,superscriptaddress,english]{revtex4-1}%
\usepackage{color}
\usepackage{amsmath}
\usepackage{amsfonts}
\usepackage{amssymb}
\usepackage{hyperref}
\usepackage{graphicx}%
\usepackage[english]{babel}

\providecommand{\U}[1]{\protect\rule{.1in}{.1in}}
\newtheorem{lemma}{Lemma}[section]
\newtheorem{theorem}[lemma]{Theorem}
\newtheorem{corollary}[lemma]{Corollary}

\def\F{{\mathbb{F}}}
\def\CHSH{{\mathrm{CHSH}}}

\begin{document}

\title{Tight bound on the classical value of $\CHSH_q(p)$ games}

\author{Matej Pivoluska}
\affiliation{Faculty of Informatics, Masaryk University,  Botanick\'a 68a, 602 00 Brno, Czech Republic}
\affiliation{Institute of Physics, Slovak Academy of Sciences, Bratislava, Slovakia}
\author{Marcin Paw\l{}owski}
\affiliation{Instytut Fizyki Teoretycznej i Astrofizyki, Uniwersytet
Gda\'nski, PL-80-952 Gda\'nsk, Poland}
\author{Martin Plesch}
\affiliation{Institute of Physics, Slovak Academy of Sciences, Bratislava, Slovakia}
\affiliation{Faculty of Informatics, Masaryk University,  Botanick\'a 68a, 602 00 Brno, Czech Republic}

\begin{abstract}
Non-local games are an important part of quantum information processing. Recently there has been an increased interest
in generalizing non-local games beyond the basic setup by considering games with multiple parties and/or with large alphabet
inputs and outputs. In this paper we consider another interesting generalization -- games with non-uniform inputs.  
Here we derive a tight upper bound for the classical winning probability for a specific family of non-local games with non-uniform
input distribution, known as  $\CHSH_q(p)$ which was  introduced recently in the context of relativistic bit-commitment protocols
by \cite{Chakraborty2015}.
\end{abstract}
\maketitle

\section{Introduction}
Non-local games are important tools of recent quantum information theory.
In a two-player non-local game a referee interacts with players who cooperate in order to win the game.
The referee chooses a pair of questions $x,y$ according to a publicly known probability distribution $r(x,y)$
and sends one question to each player. The goal of the players is to produce outputs $a$ and $b$. The win or loss
of the players is determined by a public verification function $V(a,b,x,y)\in \{0,1\}$ -- if $a$ and $b$ are valid
answers for question pair $(x,y)$, the verification function is equal to $1$ and the players win the game.
With every game $G$ we can associate two different values: the maximum winning probability of classical players
$\omega(G)$ and the maximal winning probability of players with quantum resources $\omega^*(G)$.
Finding both of these values is generally a  hard problem (in fact finding $\omega(G)$ is $NP$-hard even for
games with binary inputs and outputs
\cite{Haastad-SomeOptimalInapproximability-2001}).

Non-local games  studied in quantum information science
typically satisfy $\omega^*(G)>\omega(G)$.
This quantum advantage can be used to show various interesting results. First of all, it was originally used by John Bell
to show that there are no hidden variable extensions of quantum mechanics \cite{Bell-Einstein-Podolsky-Rosenparadox-1964}.
More recently, non-local games became central ingredients of protocols
implemented via non-communicating devices, which are able to achieve better
than classical winning probability in certain non-local games.
Their ability to exceed classical winning probability can be seen as a witness of their quantum nature.
This quantum advantage can be subsequently translated into desirable properties of the devices,
such as randomness of their outputs or monogamy of the correlations they share.
Such approach to protocol design is generally called \emph{device independence}. Examples of protocols that
can be implemented
in device independent fashion are quantum key distribution
\cite{Ac'inBrunnerGisinEtAl-Device-IndependentSecurityof-2007,VaziraniVidick-FullyDevice-IndependentQuantum-2014} and
randomness expansion and amplification  \cite{
Colbeck-QuantumAndRelativistic-2009,
PironioAc'inMassarEtAl-Randomnumberscertified-2010,
vazirani2012certifiable,
GallegoMasanesEtAl-Fullrandomnessfrom-2013,
RamanathanBrandaoHorodeckiEtAl-Randomnessamplificationagainst-2015,
BoudaPawlowskiPivoluskaEtAl-Device-independentrandomnessextraction-2014,
PleschPivoluska-Device-independentrandomnessamplification-2014,
PivoluskaPlesch-DeviceIndependentRandom-2014}.

Many communication scenarios in which non-communicating parties cooperate in order to achieve some
goal can be reduced to a non-local game. This is the reason why non-local games are also a valuable tool
in various computational complexity scenarios, such as interactive proof systems
\cite{Ben-OrGoldwasserKilianEtAl-Multi-proverInteractiveProofs:-1988}.
A recent result of Chakraborty, Chailloux and Leverrier \cite{Chakraborty2015}
is a result of this type. They were able to improve the security of a relativistic bit-commitment protocol of
Lunghi~\textit{et.~al.} \cite{Lunghi2015}
against classical adversaries into their ability to win a specific family of non-local games called $\CHSH_q(p)$.
We introduce this family of games in detail in Section \ref{sec:2}.

Chakraborty, Chailloux and Leverrier \cite{Chakraborty2015} give the following upper bound
\begin{equation}\label{eq:ChakrabortyBound}
\omega(\CHSH_q(p))\leq p + \sqrt{\frac{2}{q}}.
\end{equation}
In Section \ref{sec:4}, we derive a new upper bound for this family of games, which holds whenever
$p\geq\frac{1}{\sqrt{2q}}$
(see Theorem \ref{thm:1}
and Corollary \ref{cor}):
\begin{equation}\label{eq:OurBound}
\omega(\CHSH_q(p))\leq p + \frac{1}{2pq}.
\end{equation}
Our bound is better than the bound (\ref{eq:ChakrabortyBound}) in all instances where it holds and in fact
for certain range of parameters $q$ and $p$ it is tight as well (see Theorem \ref{thm:2}).

Our upper bound has been found by reducing the problem of finding the best classical strategy for the $\CHSH_q(p)$
games to a problem of finding the maximum amount of incidences between sets of points and sets of lines in finite fields.
This technique was  introduced by Bavarian and Shor
\cite{BavarianShor-InformationCausalitySzemeredi-Trotter-2015} in order to find upper bounds
on classical winning probability of a similar class of games called $\CHSH_q$ games. 
Later it was also used in
In Section \ref{sec:3}
we review this technique in detail.

\section{The family $\CHSH_q(p)$}\label{sec:2}
$\CHSH_q(p)$ is a family of games generalizing the well known $\CHSH$ game
\cite{ClauserHorneShimonyEtAl-ProposedExperimentto-1969}.
In the $\CHSH$ game two non-communicating players receive a single bit input $x$ and $y$ distributed
independently and  uniformly. Their goal is to provide a single bit answers  $a$ and $b$. They win the game
if $a+b = xy \mod 2$.

Recently, there has been some interest in the generalization of this game into higher alphabet
inputs and outputs
\cite{BavarianShor-InformationCausalitySzemeredi-Trotter-2015, RamanathanAugusiakMurta-XORgameswith-2015,PivoluskaPlesch-explicitclassicalstrategy-2016}.
Family of such games is called $\CHSH_q$. In these games the non-communicating players receive
uniformly distributed inputs
$x,y\in\F_q$ (where $\F_q$ denotes a finite field of size $q$) and produce outputs $a,b \in\F_q$. 
They win the game if $a+b = xy$, where addition and multiplication are both operations in $\F_q$.

Further generalization of the $\CHSH_q$ games into 
$\CHSH_q(p)$ games concerns the probability distribution of the inputs. 
$\CHSH_q(p)$ denotes a family of games with $\CHSH_q$ verification function, where Bob's input is 
distributed uniformly, while the distribution of Alice's input is independent of Bob's and is distributed 
according to some probability distribution, for which $p_\mathrm{max} \leq p$, where $p_\mathrm{max}$ 
is the probability of the her most probable input. Note that the games in this family differ only in concrete
probability distribution of Alice's input. In this paper we derive an upper bound on the classical value of these
games, which doesn't depend on this concrete distribution, but only on parameters $p$ and $q$.
With a slight abuse of notation, we call this value $\omega(\CHSH_q(p))$ and formally define it
as 
\begin{equation}\label{eq:max}
\omega(\CHSH_q(p)) = \max_i (\omega(G_i)),
\end{equation}
where the maximum is taken over all games $G_i\in\CHSH_q(p)$.

It is a well known fact that (shared) randomness does not help the classical players to win a non-local
game when compared to the deterministic strategies, since randomized strategies can be seen as convex
combination of the deterministic ones \cite{Fine-HiddenVariablesJoint-1982}. Therefore, here we consider only
deterministic classical strategies.
These can be represented by a pair of functions -- Alice's response function $a(x)$ and Bob's response
function $b(y)$. We will shorten the notation and denote the strategy as $(a,b)$.

Let us denote $\omega(q,p,r,a,b)$ the probability to win a concrete game in $\CHSH_q(p)$ 
with Alice's input distribution $r(x)$, in which players use
fixed classical strategies $a(x)$ and $b(y)$ 
It can be written as
\begin{equation}\label{eq1}
\omega(q,p,r,a,b) = \sum_{x,y\in\F_q} \frac{1}{q}r(x) V(a,b,x,y),
\end{equation}
where $V(a,b,x,y)$ is an
indicator function with value $1$ if $a(x) + b(y) = xy$. Note, that for a fixed strategy $(a,b)$ this
probability depends only on $r(x)$.
In the following lemma for fixed $q,p$ and strategy $(a,b)$ we find the best performing game from $\CHSH_q(p)$ 
-- \textit{i.e.}, we find a propability distribution $r$ maximizing $\omega(q,p,r,a,b)$.

\begin{lemma}\label{lem:1}
The probability distribution $r(x)$ maximizing
$$
\omega(q,p,r,a,b)
$$
outputs  $n-1$ elements of $\F_q$ with probability $p$,  a single  element of $\F_q$ with probability $1- p(n-1)$ and all other
elements of $\F_q$ with probability $0$, where $n = \left\lceil\frac{1}{p}\right\rceil$.
\end{lemma}

\textit{Proof.}
 Eq. (\ref{eq1}) can be rewritten as 
\begin{align}
\omega(q,p,r,a,b)  & = \frac{1}{q}\sum_{x\in\F_q} r(x) \sum_{y\in\F_q} V(a,b,x,y),
\end{align}
therefore in order to maximize this term over all 
distributions $r(x)$ we should maximize the probability of Alice's inputs $x_i\in \F_q$, which attain the highest value of
$\sum_{y\in\F_q} V(a,b,x,y)$. Since this value depends only on the response functions, we can set $r(x_i) = p$
to $n-1$ possible inputs $x_i\in \F_q$ with highest values of $\sum_{y\in\F_q} V(a,b,x,y)$.
The remaining probability ($1- (n-1)p$) can
be assigned to the input $x_i\in\F_q$ with the next highest value of $\sum_{y\in\F_q} V(a,b,x,y)$. All the other
inputs are assigned probability $0$. \hfill$\blacksquare$

The  lemma  establishes that the optimal input distribution for fixed $q$ and $p$ does not change 
(up to a permutation of elements in $\F_q$) with the response functions $(a,b)$. In other words,
the lemma identifies the game in $\CHSH_q(p)$, which obtains the maximum in Eq. \ref{eq:max}.
Let us denote this game as $\omega(\CHSH_q^{\max}(p))$.
In this game Alice's distribution of inputs assigns non-zero probability to as little values as possible.
In order to find $\omega(\CHSH_q(p))$ it remains to find an optimal strategy 
$(a,b)$ for this game, \textit{i.e.} $\omega(\CHSH_q(p)) = \omega(\CHSH_q^{\max}(p))$.

We reduce the problem  of 
problem of finding an  optimal classical strategy for $\CHSH_q^{\max}(p))$
into an instance of point-line incidence problem in $\F_q^2$. Such reduction was already successfully used by 
Shor and Bavarian \cite{BavarianShor-InformationCausalitySzemeredi-Trotter-2015} for finding upper bounds
on classical value of the $\CHSH_q$ game with uniform inputs. In the next Section we review this reduction.

\section{Point-line incidence problems and strategies for $\CHSH_q^{\max}(p)$}\label{sec:3}

A line $\ell_{a,b}\in\F_q^2$ can be characterized by two parameters -- a slope $a$ and an offset $b$.
A point $(x,y)$ lies on a line $\ell_{a,b}$, iff $y = ax+b$. Let $P \subseteq \F_q^2$ be set of points and $L \subseteq \F_q^2$
a set of lines. Let us denote $I_q(P,L)$ the amount of incidences between points $P\subseteq\F_q^2$ and lines
$L\subseteq\F_q^2$.
It is an interesting question how high $I_q(P,L)$ can be, if we fix the sizes of sets $P$ and $L$.
Indeed, this is a well studied problem in mathematics with a several important recent results \cite{Dvir-IncidenceTheoremsand-2010}.
Let us denote $I_q(n,k)$ the maximum number of point line incidences in $\F_q^2$, between set of points of size $n$ and
set of lines of size $k$.
For some pairs $(n,k)$, this optimization problem is much easier to solve than
the standard case of $n=k=q$, for which only upper and lower bounds are known \cite{Dvir-IncidenceTheoremsand-2010}.

\begin{lemma}\label{lem:2}
Let $P \subseteq\F_q^2, \vert P \vert = n$ and $L\subseteq \F_q^2, \vert L \vert  = k$. For $q \geq k \geq \frac{n(n-1)}{2}$,
\[ I_q(n,k) = k + n(n-1)/2. \] The configuration achieving this value is a complete graph with $n$ vertices,
connected with $\frac{n(n-1)}{2}$ lines (\textit{i.e.} no three points lie on the same line).
Each of the remaining lines contain exactly one point from $P$.
\end{lemma}

\textit{Proof.}
In the proof we define four different classes of configurations, which are mutually exclusive and collectively exhaustive.
Then we will show that three of these classes contain suboptimal configurations which can be improved.
We will show this by constructing a configuration with more point-line incidences for every configuration
in these classes.
The last class will contain only configurations described in the lemma and
we will show that all these strategies obtain the same (optimal) number of point-line incidences.

\begin{enumerate}
\item The first class contains all the configurations, which contain a line $\ell \in L$, without a point
from $P$.
Clearly, for all such configurations we can find a configuration with the same set of points $P$ and set of lines
$L$, in which the line $\ell$ is substituted by a line $\ell'$ that contains at least one point from $P$, achieving a larger
number of point-line incidences.

\item The  configurations in the second class do not belong to class $1$ (\textit{i.e.} each
line in $L$ contains at least one point from $P$) and there exist two points $p_1,p_2 \in P$, which are not connected by a
line from $L$. These configurations contain at most $\frac{n(n-1)}{2}-1$ lines with
more than one point from $P$. Since $L$ contains at least $\frac{n(n-1)}{2}$ lines, there exists a line in $L$ with only
a single point from $P$. We can substitute this line by a line connecting $p_1$ and $p_2$ in order to obtain a configuration
with larger number of point-line incidences.

\item The third class contains configurations which do not belong in classes $1$ and $2$ such that
$L$ contains at least one line with more than two points in $P$.

First note that for every line $\ell$ containing $c>2$ points, there are $\frac{c(c-1)}{2}-1$
lines with only a single point. This is because
$k \geq \frac{n(n-1)}{2}$, therefore $L$ contains at least as many lines as the number of pairs of points and
$\ell$ connects $\frac{c(c-1)}{2}$ pairs of points from $P$ with only a single line.

Let us now choose a point $p$ that lies on $d < n-1$ lines -- this choice together with the fact
that this configuration does not belong to class $2$ (\textit{i.~e.} $p$ is connected to every other point
by a line) ensures that $p$ shares a line with at least two other points.
In order to find a configuration with
greater number of incidences, we substitute point $p$ by a point $p'$, which does not lie on any line from $L$.
By doing this, we are decreasing the number of incidences by $d$.

Next step in creating the new improved configuration is connecting $p'$ with all other $n-1$ points
remaining in $P$. In order to do so, we need to show that the configuration of lines $L$ and points $P\backslash\{p\}$
contains at least $n-1$ lines with at most one point.
Then we will substitute these lines by lines connecting $p'$ with all points in $P\backslash\{p\}$, thus gaining
$n-1$ incidences, improving the original configuration by $n-1-d > 1$ incidences.

First we need to prove that a point not lying on any line from $L$ exists.
Note that configuration with all lines being parallel is not optimal for any values of parameters $n,k,q$.
This, together with the fact that $k\leq q$ and the fact that each line contains exactly $q$ points from $\F_q^2$,
can be used to upper bound the number of points on lines in $L$ as $kq-1 < q^2$, where $q^2$ is the number of all points
in $\F_q^2$.

In order to show that there are at least $n-1$ lines in $L$ that contain at most one point from $P\backslash\{p\}$
let us first denote $\ell_1,\dots,\ell_d \in L$ the lines going through $p$. Let us also define $c_i$ as the number of
points from $P$ the line $\ell_i$ it contains. If $c_i = 2$, then $\ell_i$ contains only a single point $p_i$
from $P\backslash\{p\}$
and we can substitute $\ell_i$ by a line connecting $p_i$ to $p'$. Let us now consider $\ell_i$ with $c_i > 2$.
As discussed earlier, for each such $\ell_i$
there are $\frac{c_i(c_i-1)}{2}-1$ lines containing only a single point in the original configuration. For
each $\ell_i$, let us denote these lines $L_i$.
We need to connect $p'$ to all points from $P\backslash\{p\}$ lying on $\ell_i$.
Since there are $c_i-1$ such points and  $\frac{c_i(c_i-1)}{2}-1 \geq c_i-1$, there are enough lines in $L_i$ to do so.

\item The fourth class contains configurations that do not belong to classes $1$, $2$ and $3$, \textit{i.e.} configurations,
in which each line contains at least one point from $P$, each two points are connected with a line from $L$
and no three points lie on the same line. These are precisely configurations defined in the lemma.
These configurations achieve the number of point line
incidences $I_q(n,k) = \frac{n(n-1)}{2} + k$. \hfill$\blacksquare$
\end{enumerate}

In order to understand how the reduction from the $\CHSH_q^{\max}(p)$ strategy to point-line incidence problem works, let us recall
the winning condition of the game $a(x) + b(y) = xy$ can be rewritten as $a(x) = xy-b(y)$.
Since $a(x)$ and $b(y)$ are Alice's and Bob's
response functions, their strategies can be seen as Alice holding a set of points $(x,a(x))$ and Bob holding a set of lines
$\ell_{y,-b(y)}$ and they give a correct answer for a pair of questions $(x,y)$, if and only if a point $(x,a(x))$ lies on a line
$\ell_{y,-b(y)}$. The only distinction to the general point line incidence problem is that Alice's set of points cannot contain
two points on the same vertical axis (otherwise there would be two possible answers for the question $x$), and
Bob's set of lines cannot contain two lines with the same slope (for the same reason). Therefore,
in order to consider a corresponding point-line incidence problem with $n$ points and $k$ lines, 
we need $q\geq n$ and $q\geq k$.

In the uniform $\CHSH_q$, both Alice and Bob have to answer all $q$ questions,
therefore we need to solve the incidence problem for $\vert P\vert = \vert L \vert = q$. Moreover, 
since all question pairs appear with probability $\frac{1}{q^2}$, the probability of a correct answer 
can be expressed as $\frac{I_q(q,q)}{q^2}$.

On the other hand, as shown by Lemma \ref{lem:1}, in $\CHSH_q^{\max}(p)$ game, Alice has to answer only
$\left\lceil\frac{1}{p}\right\rceil$ questions (other appear with probability $0$), therefore we have to solve the
point-line incidence problem for $\vert L\vert = q$ and $\vert P \vert = \left\lceil\frac{1}{p}\right\rceil = n$.
Nevertheless, this construction has to fulfill the additional constraints mentioned earlier.
Therefore the optimal value of general incidence problem shown  Lemma \ref{lem:2} is only
an upper bound for the point-line incidence reduction of an optimal strategy for $\CHSH_q^{\max}(p)$.

\section{Upper bound on classical value for selected instances of $\CHSH_q(p)$}\label{sec:4}

In this section we prove the first of the two main theorems of this paper, namely we derive a new upper 
bound on $\omega(\CHSH_q(p))$ for a certain range of parameters $q$ and $p$.

\begin{theorem}\label{thm:1}
Let $n = \left\lceil\frac{1}{p}\right\rceil$ and  $q \geq \frac{n(n-1)}{2}$. Then,
\begin{equation}\label{eqthm:1}
\omega(\CHSH_q(p)) \leq p + \frac{n-1}{q}\left(1-\frac{np}{2}\right).
\end{equation}
\end{theorem}

\textit{Proof.}
In order to prove the theorem, we use the optimal construction  from Lemma \ref{lem:2} for a problem of point line
incidences in $\F_q^2$ with $n = \vert P\vert = \left\lceil \frac{1}{p}\right\rceil$ and
$k =\vert L\vert = q$ and weight its points according to the optimal Alice's distribution introduced in Lemma \ref{lem:1}.

Without loss of generality
let us assume that all the lines in $L$ that are not part of the complete graph from Lemma \ref{lem:2}
intersect in a single point $p_1\in P$.
Therefore $p_1$ lies on precisely $q-\frac{n(n-1)}{2} + (n-1)$ lines from $L$. Since this is the point responsible for the
greatest number of incidences, it is chosen, according to the optimal distribution of Lemma \ref{lem:1}, with probability $p$.
All the other points $p_2,\dots,p_n$ lie only on lines belonging to the complete graph, therefore
each point lies exactly on $(n-1)$ lines. If the point $p_i$ is chosen,
the probability of the correct answer is given by the probability of choosing the line it lies on.
Since the lines are chosen uniformly at random with probability $\frac{1}{q}$, we get
\begin{align}
&\omega(\CHSH_q(p)) \leq \frac{q-\frac{n(n-1)}{2} + (n-1)}{q}p + (1-p)\frac{(n-1)}{q}\\
&= p -\frac{n(n-1)p}{2q}+ \frac{(n-1)p}{q} +(1-p)\frac{(n-1)}{q}\\
&= p + \frac{-n(n-1)p+2(n-1)p+2(n-1)(1-p)}{2q}\\
&=  p + \frac{n-1}{q}\left(1-\frac{np}{2}\right).
\end{align}\hfill$\blacksquare$

In the rest of this section we slightly reformulate the upper bound from Theorem \ref{thm:1}
into a form with parameters $q$ and $p$ only. This form is more practical, since it can be easily used
for comparison with the bound of \cite{Chakraborty2015} (see (\ref{eq:ChakrabortyBound})).
\begin{corollary}\label{cor}
For $p > \frac{1}{\sqrt{2q}}$,
\begin{equation}\label{eqthm:1  }
	\omega(\CHSH_q(p)) \leq p + \frac{1}{2pq}.
\end{equation}
\end{corollary}

\textit{Proof.}
First note that
\[
p + \frac{n-1}{q}\left(1-\frac{np}{2}\right) = p + \frac{(n-1)(2-np)}{2q}\leq p + \frac{1}{2pq},
\] as
$n=\left\lceil\frac{1}{p}\right\rceil$, and therefore $0\leq (n-1) \leq \frac{1}{p}$ and
$0\leq 2-np \leq 1$. Therefore we have recovered Equation (\ref{eqthm:1}).

To finish the proof it remains to show that constrain
$p\geq \frac{1}{\sqrt{2q}}$ implies $q \geq \frac{n(n-1)}{2}$.
We have:
\begin{equation}
q \geq \frac{n(n-1)}{2}
  \geq\frac{1}{2p}\left(\frac{1}{p}-1\right).
\end{equation}
In order to find out for which $p$ this inequality holds, we need to solve quadratic equation
\begin{equation}
2p^2q+p-1\geq 0
\end{equation}
for $p$, resulting in
\begin{equation}\label{eq:thmvalidity}
p \geq \frac{\sqrt{1+8q}-1}{4q} = \frac{1}{\sqrt{2q}}\left(\sqrt{1+\frac{1}{8q}}-\frac{1}{2\sqrt{2q}}\right).
\end{equation}
Therefore, for all $p\in \langle RHS,1\rangle$, where $RHS$ denotes the right hand side of equation \ref{eq:thmvalidity},
the corollary holds. To make this condition more readable, we will shorten this
interval by increasing the $RHS$.
Using $\sqrt{1+x}<1+\frac{x}{2}$, we can argue that the theorem holds whenever
\begin{equation}
p \geq\frac{1}{\sqrt{2q}}\left(1+\frac{1}{16q}-\frac{1}{2\sqrt{2q}}\right)
\end{equation}
and since $\frac{1}{16q}-\frac{1}{2\sqrt{2q}}<0$ for all $q\geq 1$, the Theorem is valid for
$\frac{1}{\sqrt{2q}}\leq p \leq 1$.
\hfill$\blacksquare$

\section{Lower bound on classical value for selected instances of $\CHSH_q(p)$}\label{sec:5}
In order to show that the bound of Theorem \ref{thm:1} is tight we need to argue that it is possible
to construct  a complete graph with $n$ vertices and $q$ edges in $\F_q^2$. This graph should also respect the additional
constraints required by
the reduction from the strategy for $\CHSH_q^{\max}(p)$ game --  no points on the same vertical line and edges 
with different slopes -- whenever $q \geq \frac{n(n-1)}{2}$.
We give such a construction for a range of $q$.

\begin{theorem}\label{thm:2}
Let $n = \left\lceil \frac{1}{p} \right\rceil$ and $q > (n-1)\left[\frac{(n-2)^2}{2}+1\right]$. Then
	\begin{equation}
		\omega(\CHSH_q(p)) \geq p + \frac{n-1}{q}\left(1-\frac{np}{2}\right).
	\end{equation}
\end{theorem}

\textit{Proof.} Here we give an algorithm to construct the set of points $P$ corresponding to Alice's 
optimal strategy for $\CHSH_q^{\max}(p)$ strategy and set of lines $L$ corresponding to Bob's 
strategy with configuration as described in Lemma \ref{lem:2}.

There are $q^2$ points in $\F_q^2$. During the run of the algorithm we maintain
two sets of points -- set $P$ of points in Alice's strategy and set $P_\mathrm{cand}$ of
candidate points that can be added into $P$ in the next round without violating the necessary conditions of a 
$\CHSH_q^{\max}(p)$ strategy.
Initialize the algorithm by $P = \emptyset$ and $P_\mathrm{cand} = \F_q^2$.
In round $i\geq 1$ of the algorithm, if $P_\mathrm{cand} \neq \emptyset$,  we move
an arbitrary point $p_i$ from $P_\mathrm{cand}$ to $P$ and then remove from $P_\mathrm{cand}$ all points
that would violate conditions of proper $\CHSH_q$ strategy:
\begin{enumerate}
\item points that lie in the same vertical line as $p_i$;
\item points that lie on lines connecting a pair of points in $P$;
\item points that lie on lines going through points of $P$ with slopes defined by some pair
of points in $P$.
\end{enumerate}
If any of these points would be added to $P$ in the next round of the protocol, they
would violate either condition for Alice's valid strategy $(1.)$ or Bob's valid strategy
$(3.)$. The condition $(2.)$ would place three points on the same line, which would violate
the optimal construction for point-line incidences.
The algorithm terminates whenever $\vert P \vert = n$, or when $P_\mathrm{cand}$ is empty and therefore no more
points can be added to $P$. After the termination the lines $L$ are defined by pairs
of points in $P$. If after the last step $\vert L\vert < q$, add to $L$ all lines with missing slopes containing point $p_1$.

In order to prove our theorem we need to show that for $q \geq (n-1)\left[\frac{(n-2)^2}{2}+1\right]$, after $n-1$ rounds of the protocol
$P_\mathrm{cand} \neq \emptyset$ therefore the $n^{\mathrm{th}}$ point can be added to $P$. In order to do this we will upper
bound the number of points
removed from $P_\mathrm{cand}$ after $n-1$ rounds.
For each of $n-1$ points in $P$, at most $q$ points in the same vertical line are removed.
Furthermore, $n-1$ points define $\frac{(n-1)(n-2)}{2}$ slopes. Together we
remove $(n-2)$ lines for each of these slopes and every line contains exactly $q$ points.
Putting this all together we remove at most $\left[(n-1) + \frac{(n-1)(n-2)^2}{2}\right]q$ points.
We require $q^2 > \left[(n-1) + \frac{(n-1)(n-2)^2}{2}\right]q$, obtaining the desired result. \hfill$\blacksquare$

\section{Conclusion}
In this paper we have introduced a new tight upper bound on the classical probability of winning games from family $\CHSH_q(p)$.
This game was introduced by \cite{Chakraborty2015} as the main tool in order 
to improve the security analysis of the relativistic bit-commitment protocol of \cite{Lunghi2015}. 
Interesting open problem is to find tight upper bounds on quantum value of these games as it might
lead to a proof of security of the said bit-commitment protocol against quantum adversaries. 
Whether our improved bounds can be directly used to improve the analysis of \cite{Chakraborty2015} remains the 
topic of our future research.

\section*{Acknowledgments}
We would like to thank Jed Kaniewski and Gl\'aucia Murta for stimulating discussions. 
Also we would like to thank Andr\'e Chailloux, whose reviews helped to shape this work
into its current form. 
MPi and MPl
were supported by the Czech Science Foundation GA\v{C}R project
P202/12/1142, EU project RAQUEL, as well as project VEGA 2/0043/15.
MPa acknowledges the support form NCN grant no. 2014/14/E/ST2/00020.

\bibliography{bitCommitment}

\end{document}